\documentclass[twocolumn,showpacs,preprintnumbers,floatfix,amsmath,amssymb,prl]{revtex4}

\usepackage{graphicx}
\usepackage{dcolumn}
\usepackage{bm}

\begin{document}
\topmargin -2cm


\title{Spin-texture and magnetic anisotropy of Co adsorbed Bi$_2$Se$_3$ topological
insulator surfaces}

\author{Tome M. Schmidt$^1$, R. H. Miwa$^1$, and A. Fazzio$^2$}
\affiliation{$^1$Instituto de F\'{\i}sica,
Universidade Federal de Uberl\^andia, Caixa Postal 593, CEP 38400-902,
Uberl\^andia, Minas Gerais, Brazil} 
\affiliation{$^2$Instituto de F\'{\i}sica, Universidade de S\~ao Paulo, 
C. P. 66318, 05315-970, S\~ao Paulo, SP, Brazil.}
\date{\today}

\begin{abstract}

Based upon first-principles methods, we investigate  the magnetic anisotropy and the spin-texture of Co adatoms embedded in the topmost Se network of the topological insulator Bi$_2$Se$_3$ surface. We find the formation of energetically stable
magnetic moment perpendicular to the surface plane, S$_z$. Our results for 
the  pristine Bi$_2$Se$_3$ surface indicate the presence of helical spin-texture not only in the massless surface Dirac states, but also surface states resonant within the valence band present spin-texture.
On the other hand, upon the presence of Co adatoms we find that the out-of-plane surface magnetism represents the dominant spin state (S$_z$),
while  the planar spin components, S$_x$ and S$_y$,  are almost suppressed.

\end{abstract}


\maketitle

Topological insulators (TIs) compose a new quantum matter phase in
solid state physics that may provide an emplacement for fundamental
physics understanding as well bases for novel technologies, such as
topological quantum computing, information processing and spintronic
applications. Recently
3D TIs were predicted by Fu et al. \cite{FuKaneMele2007} and
experimentally observed by Hsieh et al. \cite{HsiehNat2008}
in Bi$_{1-x}$Sb$_{x}$. Further, other 
compounds like Bi$_2$Te$_3$, Bi$_2$Se$_3$,
Sb$_2$Te$_3$, and TlBiSe$_2$ were identified as 3D TIs.
\cite{HsiehScie2009,HsiehNat2009,XiaNatPhys2009,ZhangNatPhys2009,
  ChenScience2009,SatoPRL2010,KurodaPRL2010} In these materials the
spin-orbit (SO) interaction rules the formation of topological states.
While the bulk is insulating, on the edges there are robust conducting 
states. Indeed, on the surface they present a (single) Dirac crossing at
the $\Gamma$ point. They show a spin-texture that allows
spin currents without dissipation, supressing back-scattering if time
reversal symmetry (TRS) is preserved. In this way the understanding of
how the protected surface states behaves in the presence of impurities
is an important subject to be explored, especially for technological
applications.  Magnetic impurities can destroy the robust surface
metalicity by breaking TRS, while nonmagnetic impurities maintain TRS.
The interaction between a
magnetic moment and the topologically nontrivial surface states 
opens a surface band gap. \cite{ChenScience2010,WrayNatPhys2011}
In this context an important issue to be investigated is 
the energetic stability of the spin orientation,
and the changes on the spin-texture upon the adsorption of transition
metal on TI surfaces, which is especially desired for spintronic and
information processing applications.

In this letter, through the calculation of magnetic anisotropy energy
(MAE) we find that Co adatoms lying on the TI Bi$_2$Se$_3$ surface
exhibit energetically stable magnetic moment perpendicular to the
surface. Further, spin-texture calculations reveal that the Co adatom
exhibits spin component $\sim$100\% polarized out-of-plane, whereas the
helical spin-texture of the massive Dirac cone is drastically reduced.
In addition, in consonance with previous experimental findings, the
presence of Co adatoms breaks TRS, opening up a surface band gap,
giving rise to massive Dirac fermions.

Our results were obtained performing {\it ab-initio} electronic structure
calculations,
using  the  density functional  theory  (DFT)  within the  generalized
gradient  approximation   (GGA)  for  the   exchange  and  correlation
potential. \cite{GGA}  The SO interaction  have been self-consistently
treated  by  using   fully  relativistic  pseudopotential  within  the
projector augmented  wave method  (PAW).\cite{PAW} We use  the  Vienna 
Ab   initio  Package  Simulation (VASP).\cite{Vasp}. A plane  wave basis
set is used,  with a cut-off energy of  300~eV.  The Brillouin-zone is
sampled,  according  to the  Monkhorst-Pack\cite{Monkhorst_PhysRevB79}
scheme, by using a number of k-points such that the total energy is
converged.
The (111)  surface of Bi$_2$Se$_3$ was investigated
using  slab method with  a vacuum  layer of  at least  8{\AA}.
Here we have used a 
thickness  of 4  quintuple  layers, keeping  the experimental  lattice
parameter. We verify that this slab is enough to obtain the correct
surface massless Dirac cone.

\begin{figure}
\includegraphics[width= 8.5cm]{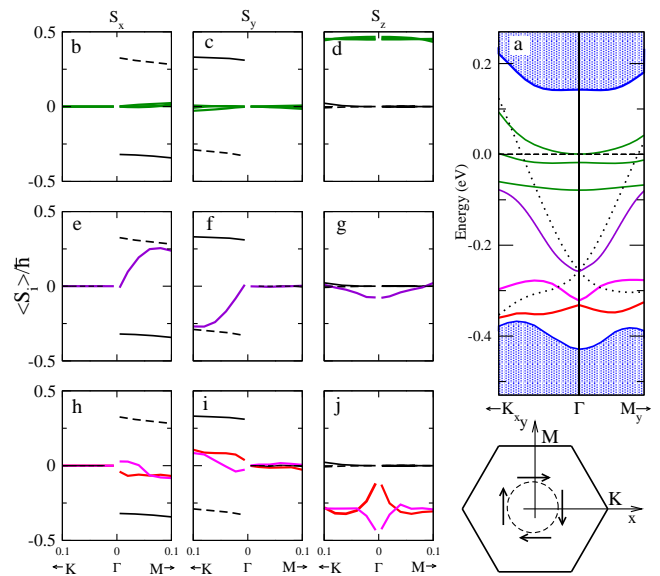}
\caption{\label{bandspin}(Color online) (a) Band structure around the bulk band
gap when a Co atom is adsorbed
on the surface of the Bi$_2$Se$_3$. The upper and lower shadow regions are the 
bulk conduction band and bulk valence band. The black dotted lines are the
original massless Dirac cones without the presence
of the Co adsorbed, and the purple line is the splitted massive Dirac cone.
The red, pink and the three green lines are the bands introduced by the Co impurity.
(b to j) Expectation values of the spin components $\langle S_{\alpha}(\vec k)\rangle$
($\alpha = x, y, z$) along the K$_x$--$\Gamma$--M$_y$ direction. (b--d) the spin of
the three upper Co-d orbitals green bands of (a); (e--g) the massive Dirac cone; 
and (h--j) are the lower Co-d orbitals (pink and red from (a)). Also
is plotted (thin black lines) the expectation values for the pristine Bi$_2$Se$_3$,
where the dotted black lines are the components for the (unoccupied) upper
massless Dirac cone and the full black lines are the components for the (occupied)
lower massless Dirac cone. At the right-bottom corner are the calculated K--$\Gamma$
and $\Gamma$--M
directions represented in the first Brillouin zone.}
\end{figure}

In  Fig.~\ref{bandspin} we show the band structure around the bulk
band gap, and the surface spin-texture when a Co atom is adsorbed
onto Bi$_2$Se$_3$ surface. The surface spin-texture 
is affected by the presence of the magnetic
impurity, as  we can see  in Fig.~\ref{bandspin}-b--j.  The original
massless Dirac cone (black dotted lines) is now splitted (purple line)
where the lower  massive Dirac cone lies  below the bulk valence
band,  and the  upper massive Dirac  cone stays inside  the
bulk band gap, Fig.~\ref{bandspin}a.
Thus, indicating that the
TRS has been removed upon  the presence of Co adatoms on the
TI surface.

The calculated expectation values of $S_x$ and
  $S_y$ of the surface protected massless Dirac cone for the
 pristine Bi$_2$Se$_3$ (black lines of Fig.~\ref{bandspin})  are
$\sim$60{\%} spin-polarized with respect to $\hbar/2$, in agreement
with a recent calculation. \cite{LuiePRL2011}
By looking at Figs.~\ref{bandspin}-e and 1-f we see that the
$S_x$ and $S_y$ for the  electronic states  of the  upper massive
  Dirac cone far from the $\Gamma$ point of the Brillouin zone, are
somewhat the
  same as compared  to the ones of  pristine Bi$_2$Se$_3$ surface.
  On  the  other hand,  for  wave  vectors  near the  $\Gamma$  point,
  $k\rightarrow\Gamma$, those spin components  tend to zero.  That is,
  the  projection of  the spin-components  on the  $xy$  plane becomes
 negligible. Meanwhile,     as      depicted     in
  Fig.~\ref{bandspin}-g, there is an increasing on the out of the plane,
  $S_z$, component  close to the  center of the Brillouin  zone, which
  can be attributed  to the formation of Co--Se  chemical bonds.  Here
  the  expectation  values  of  spin  components  are  obtained  from
$\langle  S_{\alpha}(\vec  k)\rangle  =  {{\hbar}/{2}}\langle\Psi(\vec
k)|\sigma_{\alpha}|\Psi(\vec k)\rangle$,  where $\Psi(\vec k)$  is the
spinor  wavefunction  and $\sigma_{\alpha}$  are  the Pauli  matrices.
These results  are in agreement with  recent theoretical calculations
based on few band phenomenological models.
\cite{LiuPRL2009,ZhuPRL2011,AbaninPRL2011}

We find (five) $d$ orbitals inside the bulk band gap mostly localized
in the Co atom (Fig.~\ref{bandspin}-a). We can separate
these orbitals  in two  classes according to  their symmetry.   Two of
them  (red and  pink lines)  suffer  a strong
hybridization with  the
surface  Se atoms,  resulting  in a  quasi-Dirac  crossing at the
$\Gamma$ point. These  two
orbitals  have Se-$p_z$  and Co-$d_{xy}$  and  -$d_{x^2-y^2}$ symmetry
characters. As we can see from Fig.~\ref{bandspin}-h  and \ref{bandspin}-i
they present small  $S_x$ and $S_y$  spin components, and
  the left-handed  spin helicity has  been maintained,
  while there  is an out  of plane $S_z$  of around $\hbar/4$  for the
  electronic   states   near  the   edge   of   the  Brillouin   zone,
  Fig.~\ref{bandspin}-j. The other
three  Co-$3d$ orbitals  (green lines  of the  Fig.~\ref{bandspin}-a) are
more  localized,  being  two  occupied  bands  ($d_{xy-yz}$)  and  one
unoccupied  ($d_{z^2}$).  The magnitude of the $S_z$  component of these
three  bands are $\sim\hbar/2$ spin
aligned  perpendicular  to  the   TI  surface  (see
Fig.~\ref{bandspin}-d), and do not dependent on
k-direction. In contrast the Co-$d_{xy}$  and  -$d_{x^2-y^2}$
(Fig.~\ref{bandspin}-j), due to the hybridization, the spin
polarization is dependent on the k-direction.

\begin{figure}[h b t]
\includegraphics[width= 3cm]{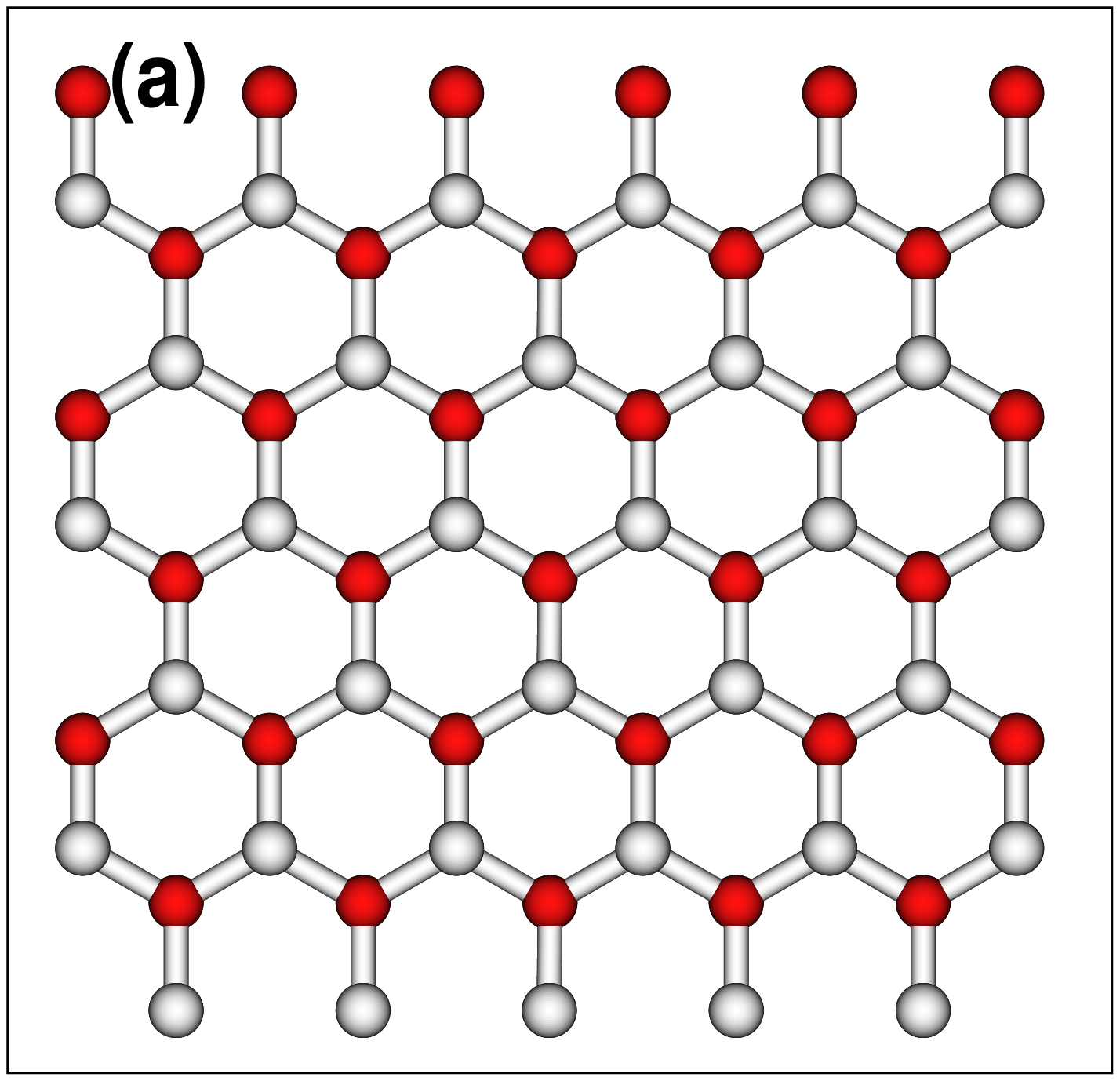}
\includegraphics[width= 3cm]{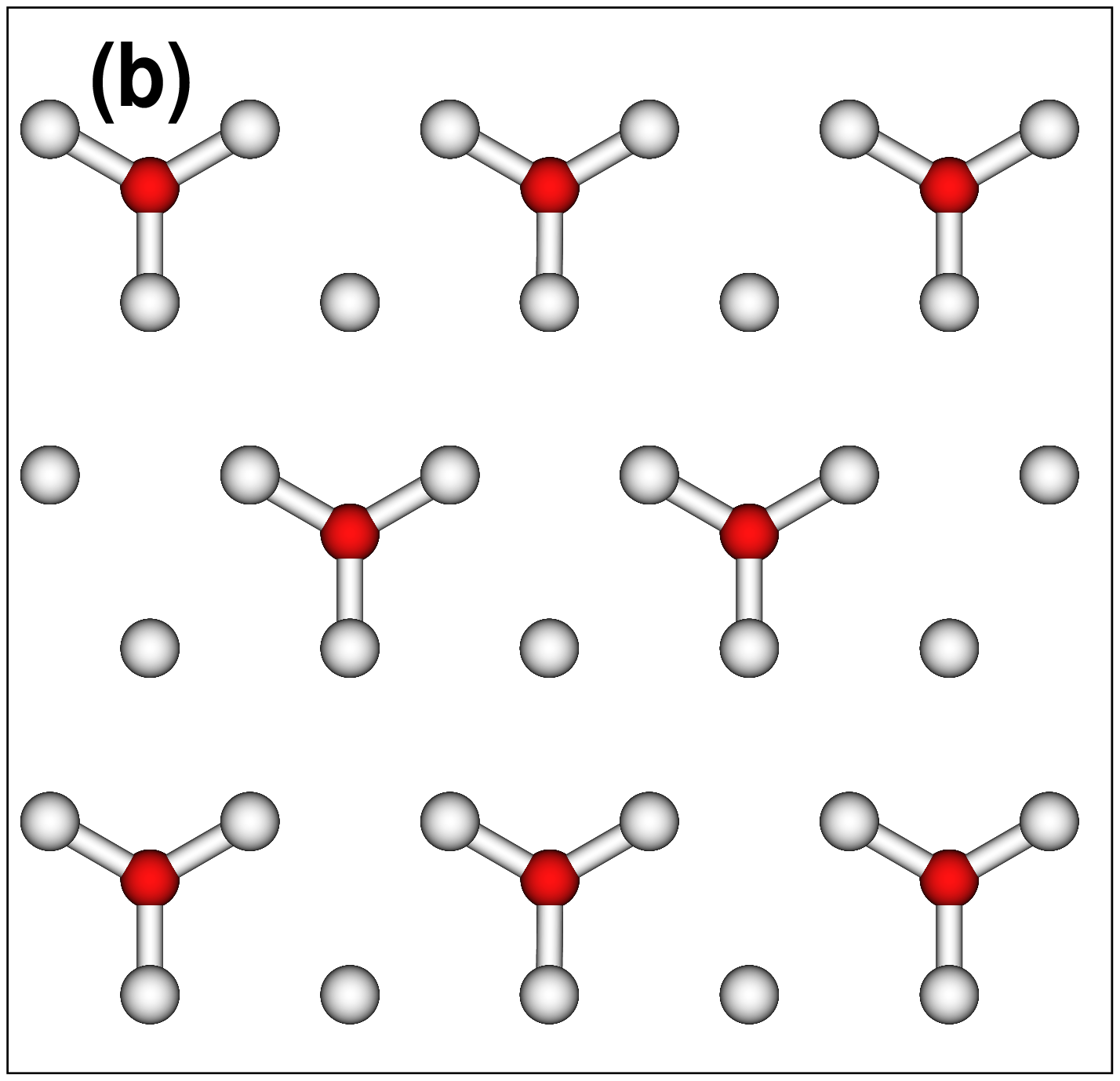}
\includegraphics[width= 3cm]{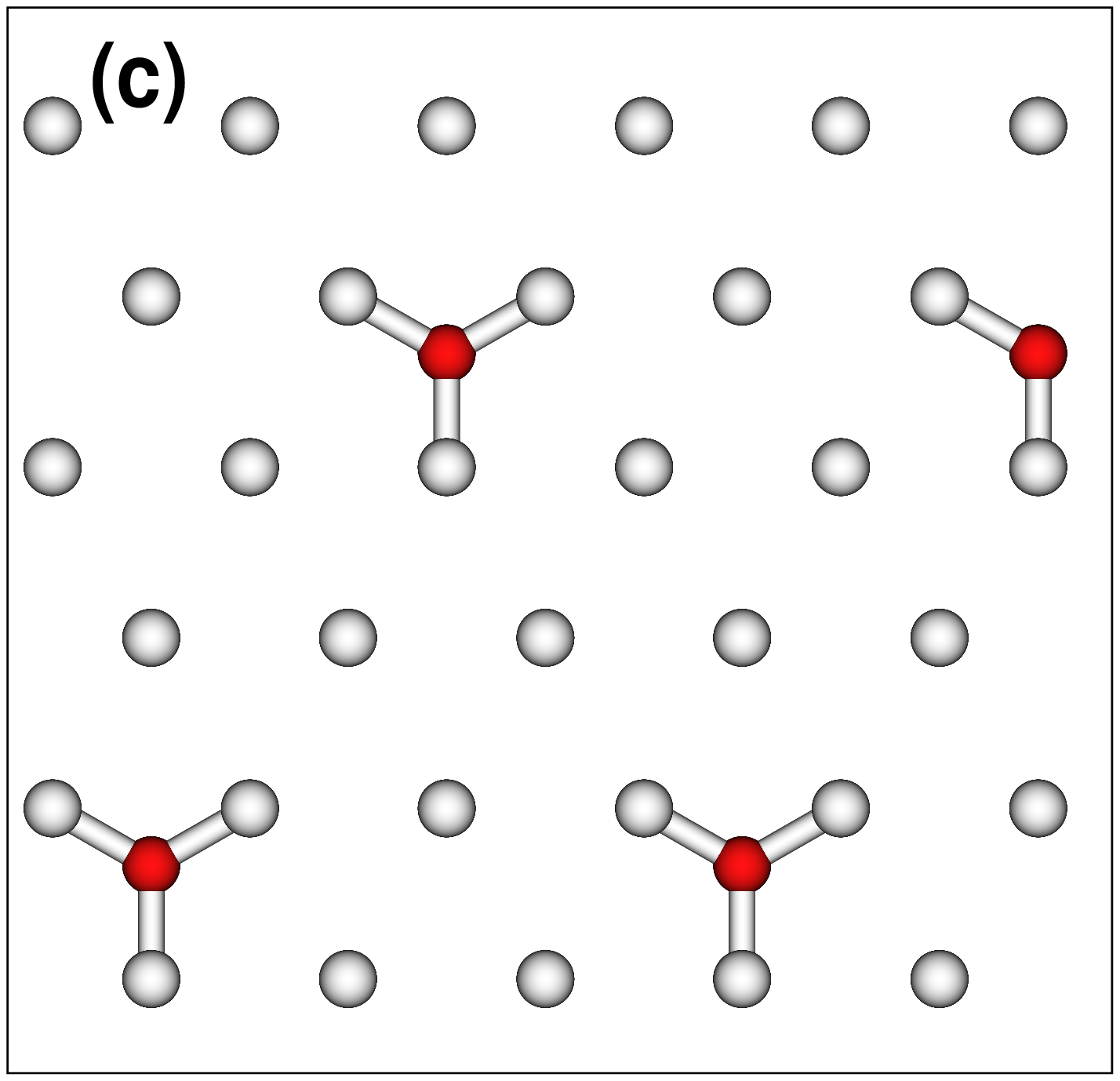}
\includegraphics[width= 3cm]{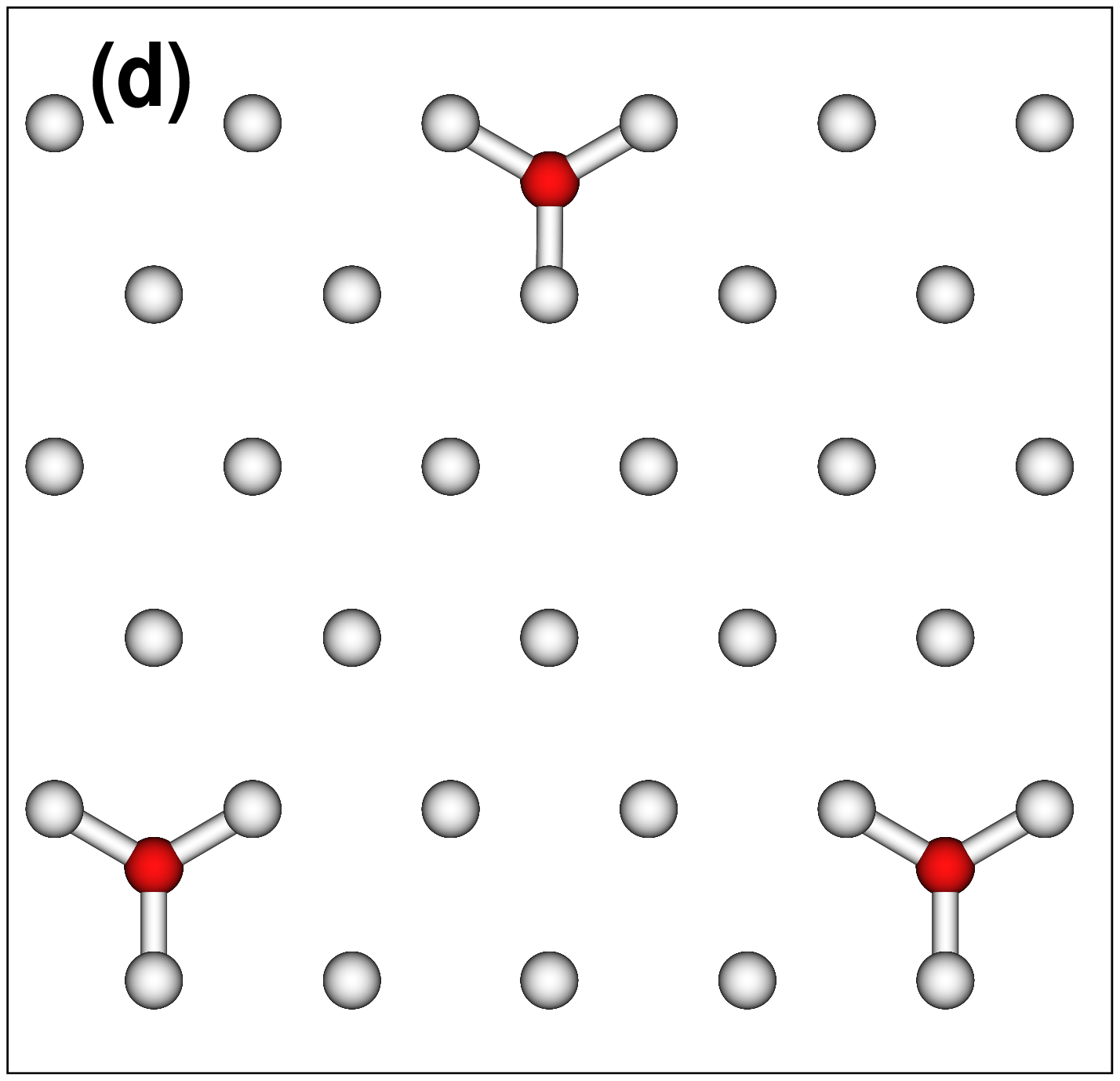}
\caption{\label{struct} (Color online) Schematic representantion of Co
adsorbed on the TI
surface for 1, 1/4, 1/9 and 1/16 monolayer coverages, from (a) to (d),
respectively. The whiter balls represent the Se topmost surface layer, and the
red balls represent the embedded Co atoms.}
\end{figure}

It is worth to point out that the results presented in Fig.~\ref{bandspin}
are for a Co coverage of 1/9 monolayer (ML). In addition we have considered
coverages of 1, 1/4 and 1/16~ML, as schematicly depicted in Fig.~\ref{struct}.
For 1/9 ML Co coverage on the  surface of Bi$_2$Se$_3$ there is not new Dirac
point, but
two  bands (red  and pink lines of this  figure) are  getting close  at the
$\Gamma$ point. By  increasing the Co coverage to 1 or 1/4 ML, this two bands
are splitted. However by  decreasing the Co coverage  to 1/16 ML
the two bands touch to each  other at the $\Gamma$ point, forming a new
Dirac cone,  similarly as experimentally observed  for Fe deposited 
on Bi$_2$Se$_3$.
\cite{HasanNatPhys2011}  We verify that the spin-texture for the other
bands (then the red and pink ones from Fig.~\ref{bandspin}) are similar
for all Co coverages.

\begin{figure}
\includegraphics[width= 7.5cm]{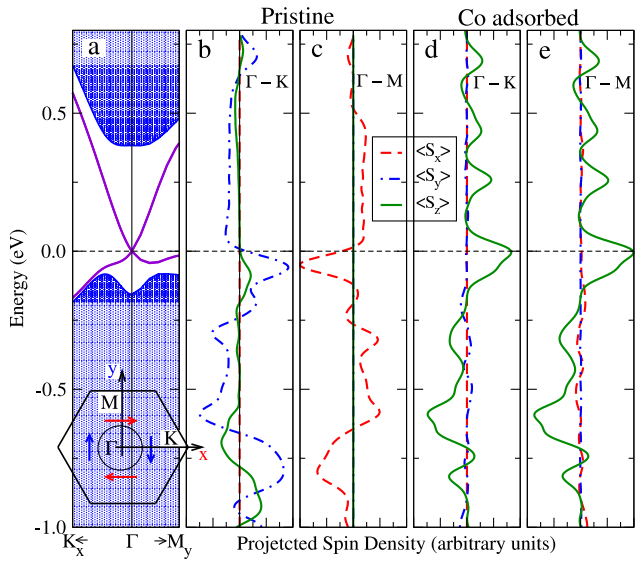}
\caption{\label{SpinDos} (Color online) Band structure (a) and
Spin-Density projected on each S$_{\sigma}$ spin component for
pristine Bi$_2$Se$_3$ along the $\Gamma$-K (b) and $\Gamma$-M (c) directions.
(d--e) The Spin-Density projections for the system when a Co atom is
adsorbed on the TI surface.
The positive and negative signs for $<$S$_{\sigma}>$ are with respect to the
K$_x$--$\Gamma$--M$_y$ directions, as indicated inset at the hexagon diagram.
The left-handed helicity shown in the hexagon diagram is for a cut above the
Fermi level.}
\end{figure}

In order to understand the spin projection as a function of the energy
states we plot  in Fig.~\ref{SpinDos} the Spin-Density 
projected  for each  spin component  ($\alpha =  x,y,z$). This projection
was computed  for  a mesh  of  k-points  along the  K$_x$--$\Gamma$--M$_y$
direction. For pristine  Bi$_2$Se$_3$ (Fig.~\ref{SpinDos}-b and -c) we
observe that  the occupied massless  Dirac  cone  presents  higher 
density  of  spins  than  the unoccupied one.
From this figure we also observe that there is a spin-texture
for states inside the bulk valence band reaching $\sim$1~eV below the Fermi
level. These spin-textures are due to the presence of surface
states resonant within the bulk valence band.
When the Co atom is  adsorbed on the TI surface (Fig.~\ref{SpinDos}-d and
-e) the  $\langle S_x\rangle$ and  $\langle S_y\rangle$  components 
reduce drastically,
while   the   $\langle  S_z\rangle$   increases,   showing  that   the
out-of-plane  surface magnetism  are the  dominant spin  states.

\begin{figure}
\includegraphics[width= 8cm]{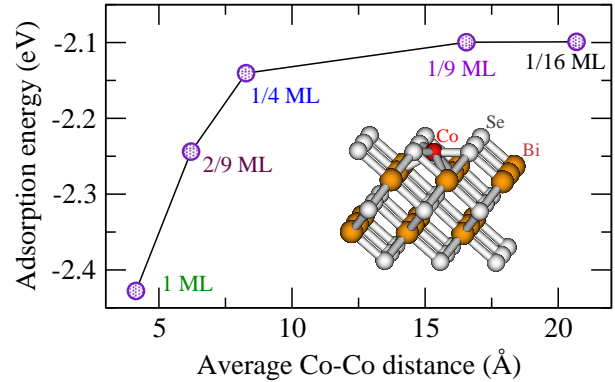}
\caption{\label{adsorption} (Color online) Adsorption energy as a function of Co-Co average
distance distributed on the surface of Bi$_2$Se$_3$. The adsorption energy
is computed with respect to the clean surface and isolated Co$_2$ dimer.
Inset is the atomic arrangement when a Co atom is adsorbed on the Bi$_2$Se$_3$
surface. The Co atom  is bonded to the first (Se) and second (Bi) atomic layers.}
\end{figure}

The Co atom adsorbed on the Bi$_2$Se$_3$ surface is bonded to three top surface
Se and three Bi atoms of the second layer, as showed inset of
Fig.~\ref{adsorption}.
This system is exothermic by 2.09~eV per Co atom with respect to clean surface and
isolated Co dimers. There is an attractive interaction between the Co atoms when
they are adsorbed on the Bi$_2$Se$_3$ surface. By increasing the Co coverage,
as we can see in Fig.~\ref{adsorption} the adsorption energy increases. The
Co-Co interaction favors a FM state. The total energy difference per pair of Co
atoms with spin parallel as compared to that of spin anti-parallel is 57~meV
favoring a FM configuration. A FM phase was theoretically predicted for magnetic
impurities in TI when S$_z$ spin component is dominant. 
\cite{LiuPRL2009,AbaninPRL2011} This out-of-plane spin net can works as a local
magnetic field inducing an anomalous Hall effect. \cite{ZhuPRL2011,YuScience2010}

\begin{figure}[h,t,b]
\includegraphics[width= 8cm]{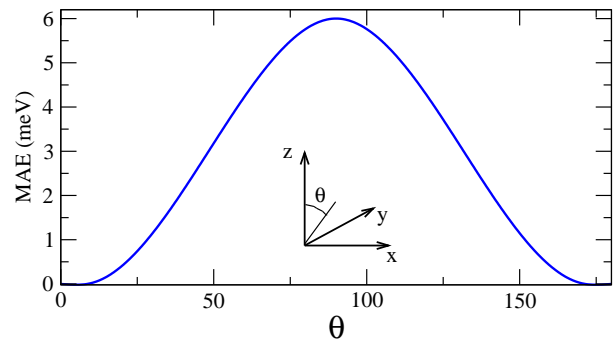}
\caption{\label{MAE} Magnetic anisotropy energy
computed as the total energy as a function of the magnetic moment alignment. The  zero
angle is for a magnetic moment aligned out-of-plane, perpendicular to the TI surface
($z$ direction). $xy$ is the basal TI plane.}
\end{figure}

The spin-texture of the TI--magnetic-impurity system is dependent on the
spin polarization.  In order to verify the energetic stability of our
calculated spin-texture (Fig.~\ref{bandspin} and \ref{SpinDos}) we
calculate the magnetic anisotropy energy of the magnetization induced
by the Co adatoms on the Bi$_2$Se$_3$ surface. As we can see in
Fig.~\ref{MAE} the minimum energy occurs for the magnetic moment aligned
perpendicular to the TI surface, and the energy barrier is verified when the
magnetic moment is in-plane. A single Co impurity adsorbed on TI surface
presents a MAE of 6~meV per impurity.

In summary our first principles calculations provide a fundamental
understanding of TI properties upon the adsorption of a magnetic impurity.
The adsorbed transition metal Co impurity,
give rise to massive Dirac fermions,
inducing an energetically stable out-of-plane spin component, S$_z$,
ruled  by the Co 3$d$ electronic states lying within the bulk band gap.
This S$_z$ component is $\sim$100\% polarized, whereas the polarization
of the surface
helical spin-texture of the massive Dirac cone is drastically reduced with
respect to pristine Bi$_2$Se$_3$.
The impurity introduces a net spin with MAE of
6~meV per adsorbed Co atom, favoring the magnetic moment to be
aligned perpendicular to the TI surface.
These results have a direct connection with information processing and
spintronic applications using topological devices.

\begin{acknowledgments}
This work was supported by the Brazilian agencies FAPEMIG, FAPESP, CAPES, CNPq, and
the computational support from CESUP/UFRGS.
\end{acknowledgments}

\end{document}